\documentclass[aps,a4paper]{revtex4}

\usepackage{amssymb} \usepackage{amsmath,amssymb,graphicx,color,epsfig}
\usepackage{pstricks,float}
\usepackage[toc,page]{appendix}
\usepackage{subfig, tensor}
\usepackage{graphicx}   
\usepackage{caption}    
\usepackage{subcaption} 
\usepackage{subcaption}
\begin{document}
\title{Warm Hybrid Axion Inflation in $\alpha$-Attractor Models Constrained by  ACT and Future Plan experiments.}
\author{Waqas Ahmed$^{1}$\footnote{E-mail: \texttt{\href{mailto:waqasmit@hbpu.edu.cn}{waqasmit@hbpu.edu.cn}}}, Waqar Ahmad$^{2,4}$\footnote{E-mail: \texttt{\href{mailto:waqarahmad262@gmail.com}{waqarahmad262@gmail.com}}},
Ahsan Illahi$^{3}$\footnote{E-mail: \texttt{\href{mailto:ahsanillahi@comsats.edu.pk}{ahsanillahi@comsats.edu.pk}}}
M. Junaid$^{4}$\footnote{E-mail: \texttt{\href{mailto:mjunaid@ualberta.ca}{mjunaid@ualberta.ca}}} 
}

\affiliation{$^1$ Center for Fundamental Physics, School of Artificial Intelligence, Hubei Polytechnic University, Huangshi 435003, China.\\
$^2$ School of Natural Sciences, Department of Physics and Astronomy, National University of Sciences and Technology, Islamabad, Pakistan.\\
$^3$ Department of Physics, COMSATS University Islamabad, Islamabad, Pakistan.\\
$^4$National Centre for Physics, Islamabad, Pakistan.}


\begin{abstract}
We present a comprehensive study of warm hybrid inflation within the framework of $\alpha$-attractor models, where an axionic inflaton is coupled to a waterfall field in the presence of thermal dissipation. The model is analyzed for both linear ($\Upsilon \propto T$) and cubic ($\Upsilon \propto T^{3}$) dissipation regimes. Confronting the theoretical predictions with the latest observational data from Planck+BICEP/Keck, P-ACT-LB-BK18 and  SPT, and , we find that in the weak dissipative regime ($Q_{*} \lesssim 10^{-5}$), the scalar spectral index $n_{s} \simeq 0.965$ lies at the boundary of the combined P-ACT-LB-BK18 constraints, while the tensor-to-scalar ratio $r$ remains within observable ranges. For stronger dissipation ($Q_{*} \gtrsim 10^{-5}$), the model predicts values of $n_{s}$ well within the $1$--$2\sigma$ confidence region of all datasets, with tensor modes remaining fully observable in both dissipation scenarios. These results indicate that forthcoming CMB polarization experiments may be capable of detecting primordial gravitational waves, thereby providing a robust observational test of warm hybrid inflation across different dissipative regimes.

\end{abstract}

\maketitle

\section{Introduction}

Inflationary cosmology stands as the cornerstone of modern earl universe theory, providing a robust, predictive framework that explains the remarkable homogeneity, isotropy, and flatness of the observable universe\cite{Guth:1980zm, Linde:1981mu}. Inflation provides a beautiful solution to the traditional horizon and flatness problems of the Hot Big Bang model by proposing a phase of faster, quasi-exponential growth fuelled by the potential energy of an inflaton field, a scalar field. Most importantly, it also provides a quantum-mechanical mechanism by which the primordial density perturbations which flattened the cosmic microwave background (CMB) anisotropies and the universe in general were seeded to occur\cite{Mukhanov:1981xt}.

The success of inflation, however, is tempered by a fundamental theoretical challenge: the transition from a compelling cosmological paradigm to a concrete model grounded in particle physics. The central obstacle is the so-called eta-problem: maintaining the requisite flatness of the inflaton potential over super-Planckian field ranges while protecting it from large quantum corrections that would otherwise spoil the slow-roll conditions \cite{Easson:2009kk, McAllister:2007bg}. This problem is exacerbated by the swampland conjectures in quantum gravity, which question the viability of sustained slow-roll inflation in regimes controlled by effective field theories weakly coupled to gravity \cite{Obied:2018sgi, Agrawal:2018own}.

Traditionally, model building has sought refuge in symmetries such as supersymmetry or shift symmetries of pseudo-Nambu Goldstone bosons (axions) to protect the inflationary potential~\cite{Freese:1990rb,Silverstein:2008sg}. However, these approaches often introduce their own fine-tuning problems or come into tension with observational bounds and theoretical constraints, such as the Weak Gravity Conjecture~\cite{Rudelius:2015xta,Stein:2021uge}. Furthermore, recent advances in precision cosmology have rendered many simple and well-motivated particle physics models including monomial chaotic inflation scenarios with potentials $V(\phi)\propto \phi^{2}$ or $V(\phi)\propto \phi^{4}$, as well as attractor-type models such as Starobinsky and Higgs inflation---increasingly disfavored by observational data~\cite{Kallosh:2025rni,AtacamaCosmologyTelescope:2025nti}.

This tension is starkly highlighted by the evolving observational landscape. Early data from the \emph{Planck} satellite, combined with constraints from BICEP/Keck (BK), pointed to a scalar spectral index of
\[
n_s \simeq 0.9652 \pm 0.0084 \quad (95\%~\mathrm{CL}) ,
\]
as reported in Refs.~\cite{Planck:2018jri,BICEP:2021xfz}. However, more recent multi-experiment analyses have shifted the preferred value of $n_s$ toward higher values. In particular, the so-called P-ACT-LB-BK18 dataset combining Data Release~6 from the Atacama Cosmology Telescope (ACT), \emph{Planck}, BICEP/Keck, and DESI baryon acoustic oscillation measurements yields
\[
n_s = 0.9743 \pm 0.0068 \quad (95\%~\mathrm{CL}:~0.967 \lesssim n_s \lesssim 0.981) ,
\]
as shown in Refs.~\cite{AtacamaCosmologyTelescope:2025nti,AtacamaCosmologyTelescope:2025blo,DESI:2024mwx}. Meanwhile, an independent combination of South Pole Telescope (SPT), \emph{Planck}, and ACT data (P-ACT-SPT) favors
\[
n_s = 0.9684 \pm 0.006 \quad (95\%~\mathrm{CL}:~0.962 \lesssim n_s \lesssim 0.974) ,
\]
as reported in Ref.~\cite{SPT-3G:2025vtb}. This upward trend in $n_s$, together with increasingly stringent upper limits on the tensor-to-scalar ratio,
\[
r < 0.036 \quad (95\%~\mathrm{CL}),
\]
This poses a significant challenge to many canonical ``cold'' inflationary scenarios. In this context, numerous studies have recently investigated cold inflationary models, both within supersymmetric and non-supersymmetric frameworks, in light of observational constraints from the Atacama Cosmology Telescope (ACT); see Refs.~\cite{Rehman:2025fja, Kallosh:2025rni, Zharov:2025zjg, Ketov:2025cqg, Pallis:2025epn, Okada:2025nyd, Pallis:2025nrv, Pallis:2025gii, Pallis:2025vxo, Ellis:2025zrf, Gao:2025onc, Liu:2025qca, Odintsov:2025eiv, Gialamas:2025kef, McDonald:2025tfp, Modak:2025bjv, Ahmed:2025eip}. These developments in cold inflation, particularly in explaining the P-ACT-LB-BK18 and SPT observations, also motivate us to explore axion hybrid inflation within the framework of warm inflation, which is itself well motivated, and to examine these results in this broader context
~\cite{Berera:2025vsu, Berera:2023liv}.

A promising and theoretically well-motivated path forward is provided by the paradigm of \emph{warm inflation} (WI)~\cite{Berera:1995ie,Berera:1999ws}. 
In contrast to standard ``cold'' inflation (CI), where the inflaton is assumed to be dynamically isolated from other degrees of freedom until a post-inflationary reheating phase, WI posits that sizable dissipative interactions between the inflaton and a thermal bath are already active during the inflationary era. 
These interactions are characterized by a dissipation coefficient $\Upsilon$, which introduces an additional friction term $\Upsilon \dot{\phi}$ in the inflaton equation of motion. 
As a result, energy is continuously transferred from the inflaton to radiation, maintaining a non-negligible radiation energy density $\rho_r$ throughout inflation.

This simple yet profound physical ingredient leads to several important consequences. 
First, it alleviates the $\eta$-problem: in the strong dissipative regime,
\begin{equation}
Q \equiv \frac{\Upsilon}{3H} \gg 1 ,
\end{equation}
slow-roll evolution can be sustained even if quantum corrections generate an inflaton mass $m_\phi$ larger than the Hubble scale $H$, a situation that is incompatible with CI~\cite{Berera:2003yyp}. 
Second, WI naturally yields a smooth and graceful exit into a radiation-dominated universe without requiring a separate reheating phase~\cite{Berera:1996fm}. 
Third, and most relevant for current observations, the dissipative dynamics and associated thermal fluctuations significantly modify the primordial power spectrum.

In warm inflation, the scalar curvature power spectrum receives enhanced contributions from thermal noise. 
It can be written as \cite{Ramos:2013nsa}
\begin{equation}
\mathcal{P}_{\mathcal{R}}(k)
=
\left(
\frac{H^2}{2\pi \dot{\phi}}
\right)^2
F ,
\end{equation}
where the enhancement factor $F$ is given by
\begin{equation}
F
=
\left(
1 + 2 n_* +
\frac{2\sqrt{3}\pi Q}{\sqrt{3+4\pi Q}}
\frac{T}{H}
\right)
G(Q) .
\end{equation}
Here, $n_*$ accounts for the possible thermalization of inflaton fluctuations, while $G(Q)$ encodes the growth of perturbations due to the coupling between inflaton and radiation fluctuations and depends on the microphysical structure of the dissipation coefficient $\Upsilon$~\cite{Berera:2025vsu}.

These modifications can naturally yield a scalar spectral index $n_s$ closer to unity and suppress the tensor-to-scalar ratio $r$, in excellent agreement with the trends favored by recent P-ACT-LB-BK18 and SPT observations~\cite{AtacamaCosmologyTelescope:2025nti}.

Recent advances in numerical analysis, particularly through dedicated codes such as \texttt{WI2easy}~\cite{Rodrigues:2025neh}, now enable precision calculations of both the background dynamics and cosmological perturbations in unified warm little inflaton (WLI) models. 
These studies demonstrate that the strong dissipative regime,
\begin{equation}
Q \gg 1 ,
\end{equation}
of the combined framework is not only theoretically well motivated—ensuring sub-Planckian field excursions,
$\Delta \phi < M_{\mathrm{Pl}}$, and avoiding potential swampland constraints~\cite{Motaharfar:2018zyb,Berera:2019zdd} but is also explicitly favored by the preference for a higher scalar spectral index $n_s$ indicated by the combined P--ACT--LB--BK18 data sets~\cite{AtacamaCosmologyTelescope:2025blo,AtacamaCosmologyTelescope:2025nti}. 
Moreover, warm inflation naturally accommodates a mildly positive running of the scalar spectral index,
\begin{equation}
\alpha_s > 0 ,
\end{equation}
as suggested by recent observations, a feature that remains challenging to realize within many cold inflation scenarios.

Motivated by these results, the present work aims to explore a related yet complementary direction: the incorporation of warm inflationary dynamics within the framework of hybrid inflation~\cite{Linde:1993cn, Ahmed:2024tlw}. 
Conventional hybrid inflation models, which employ a two-field mechanism to terminate inflation, typically neglect the potential role of sustained thermal effects during the inflationary phase. By combining the warm inflation paradigm with a hybrid structure with axionic background we seek to construct a robust and microphysically consistent inflationary scenario.

Our objective is to elucidate how the interplay between dissipative thermal friction, multi-field dynamics influences the inflationary observables, the reheating dynamics, and the connection to fundamental energy scales. 
To this end, we perform both analytical and numerical explorations of the model’s parameter space, confronting its predictions for the scalar spectral index $n_s$, the tensor-to-scalar ratio $r$ and the running $\alpha_s$ with the stringent constraints arising from the combined analyses of \emph{Planck+BICEP/Keck}, P-ACT-LB-BK18 and SPT data. In doing so, we aim to contribute toward building a principled and observationally viable bridge between inflationary cosmology and high-energy fundamental physics.

The rest of the work is structured in the following manner. In Section II we introduce the theoretical framework of warm hybrid axion inflation in the framework of the $\alpha$-attractor formalism, including the two-field potential, axionic self-interactions, and the warm inflation background equations. Section III outlines the dynamics of warm inflation, dissipation regimes, and the analytical formulation for cosmological perturbations. Section~IV contains our comprehensive numerical analysis, exploring the full parameter space and presenting results for $n_s$, $r$, $\alpha_s$, and $T_*/H_*$ as functions of the dissipation ratio $Q_*$, along with a comparison to current observational datasets.  Finally, in Section~V, we summarize the key findings and conclude the work.

\section{Hybrid Inflation with Axion Sector}

We investigate a hybrid warm inflation scenario involving axions (or axion-like particles), where the scalar field $\phi$ plays the role of the inflaton while $\psi$ acts as the waterfall field. Unlike the standard cold picture, here the inflaton evolves in the presence of a thermal radiation bath maintained through dissipative processes. Additionally, we consider a non-canonical modification of the kinetic term of $\phi$ within the framework of an $\alpha$-attractor model. The resulting Lagrangian density is \cite{Braglia:2022phb}

\begin{equation}
L \simeq
\dfrac{(\partial^\mu\phi)^2}{2\left(1-\dfrac{\phi^2}{6\alpha}\right)^2}
+\dfrac{(\partial^\mu\psi)^2}{2}
-V(\phi,\psi),
\end{equation}

where warm inflation effects are introduced through a dissipation coefficient $\Upsilon(\phi,T)$ that transfers inflaton energy to a thermal bath which we discuss in detail next section. The background evolution equations are modified as
\begin{align}
\ddot{\phi} + (3H + \Upsilon)\dot{\phi} + V_{\phi} &= 0, \quad
\dot{\rho}_r + 4H\rho_r = \Upsilon \dot{\phi}^2,
\end{align}
where $\rho_r$ is the radiation energy density.

The scalar potential is chosen as
\begin{equation}
V(\phi,\psi)=
\kappa^2\left(M^2-\dfrac{\psi^2}{4}\right)^2
+V(\phi)
+\dfrac{\lambda^2}{4}\phi^2\psi^2,
\end{equation}
with the axionic self-interaction given by
\begin{equation}
V(\phi)=
f^2 m^2\left[1-\cos\left(\dfrac{\phi}{f}\right)\right].
\end{equation}

Here $M$ and $m$ denote mass scales associated with the waterfall and inflaton sectors, respectively, while $f$ is the axion decay constant and $\kappa,\lambda$ are dimensionless couplings. In the warm inflation regime, the presence of dissipation slows the evolution of $\phi$ near the critical point, causing the tachyonic instability of $\psi$ to grow gradually rather than abruptly. 

The parameter $\alpha$ originates from the hyperbolic geometry of $\alpha$-attractor models \cite{Kallosh:2025ijd,Galante:2014ifa}. The associated kinetic term exhibits a pole at $\phi=\sqrt{6\alpha}$, stretching the potential in the canonical field frame and modifying inflationary observables such as $n_s$ and $r$. Similar kinetic structures can also arise from renormalization group effects in Higgs sector extensions \cite{Ghoshal:2023pcx}. Introducing the canonical inflaton field via
\begin{equation}
\phi \rightarrow \sqrt{6\alpha},\tanh\left(\dfrac{\varphi}{\sqrt{6\alpha}}\right),
\end{equation}
the hybrid potential becomes
\begin{align}
\label{canonpoten}
V(\psi,\varphi) &= \kappa^2\left(M^2-\dfrac{\psi^2}{4}\right)^2+m^2f^2\left[1-\cos\left(\frac{\sqrt{6\,\alpha}\, \text{Tanh}\left(\dfrac{\varphi}{\sqrt{6\,\alpha}}\right)}{f}\right)\right] \\\notag
&+\dfrac{\lambda^2}{2}\psi^2\left(\sqrt{6\,\alpha}\, \text{Tanh}\left(\dfrac{\varphi}{\sqrt{6\,\alpha}}\right)\right)^2 .
\end{align}
 The mass squared of the waterfall field at $\psi=0$ is,
\begin{align}
\label{waterfallmass}
M_\psi^2=\left(-\kappa^2\,M^2+\dfrac{1}{2}\left(\lambda\,\sqrt{6\,\alpha}\,\text{Tanh}\left(\dfrac{\varphi}{\sqrt{6\,\alpha}}\right)\right)^2\right).
\end{align}

To ensure stability along $\psi=0$ during inflation, we assume
\begin{equation}
\dfrac{(\lambda\sqrt{6\alpha})^2}{2} > \kappa^2 M^2,
\end{equation}
so that $M_\psi^2 > 0$ for $\varphi>\varphi_c$. The critical condition is
\begin{align}
    \text{Tanh}^2\left(\dfrac{\varphi_c}{\sqrt{6\,\alpha}}\right)=\dfrac{V_{0}^{1/2}}{3\,\alpha\,\lambda_c}.
\end{align}
where $V_0 = \kappa^2M^4$ and $\lambda_c = \lambda^2/\kappa$. Along the inflationary trajectory $\psi=0$, the effective single-field potential becomes
\begin{equation}
V(\varphi) = V_0 + m^2f^2\left[1-\cos\left(
\dfrac{\sqrt{6\alpha}\tanh\left(\dfrac{\varphi}{\sqrt{6\alpha}}\right)}{f}
\right)
\right],
\end{equation}
which drives the warm inflation dynamics together with the thermal bath. The resulting expressions for the scalar power spectrum, tensor-to-scalar ratio, and their dependence on the dissipation strength will be analyzed in subsequent sections.

\section{Warm Inflation}

Warm inflation is the realization that the interactions between the inflaton and other fields as radiation can lead to dissipation of inflaton energy to other dynamical degrees of freedom. This means that particle production can occur concurrently with inflationary expansion as long as the scalar potential remains the dominant component of energy density in the universe, with the ambient temperature greater than the Hubble scale. Radiation can naturally come to dominate the energy density of the universe through this particle production without the need for the separate reheating phase of cold inflation\cite{Berera:2023liv,Benetti:2016jhf}.


In the warm inflation scenario, inflaton interacts with other field(s) which is characterized by a dissipation coefficient. The evolution equation to govern the dynamics of warm inflaton field is
\begin{equation}
\ddot{\phi} +(3H+\Upsilon)\dot{\phi} + V_{,\phi}=0
\label{EvolutionInflatonEq}
\end{equation}
where $\Upsilon$ is a dissipation coefficient, which is absent in the cold inflation. Whereas, the evolution of energy density can be extracted by invoking the energy conservation which reads as
\begin{equation}
 \dot{\rho}_r+4H\rho_r= \Upsilon \dot{\phi}^2
\label{EvolutionEnergyEq}
\end{equation}
The term $\Upsilon \dot{\phi}^2$ encodes the transfer of energy from the inflaton and the radiation bath. In warm inflation, it is assumes that the radiation to be thermalized:
\begin{equation}
    \rho_r=(\pi^2/30)g^* T^4
\end{equation}
where $g^*$ signifies the light degrees of freedom for the radiation bath. Finally, the Friedmann equation takes the following form:
\begin{equation}
    H^2 = \frac{1}{3m_p^2}\left(\frac{\dot{\phi}^2}{2}+V(\phi) + \rho_r \right)\label{HubbleSqr}
\end{equation}
where $m_p=1/\sqrt{8\pi G}$ is the reduced Planck mass.
\\
In warm inflation slow roll approximations remain applicable. In slow-roll regime, where the 2nd order derivative of inflaton field, $\phi$ in Eq.~\ref{EvolutionInflatonEq} and 1st order derivative of energy density, $\rho_r$, in Eq.~\ref{EvolutionEnergyEq} can be neglected, the evolution equations for inflaton and energy densities turn to be
\begin{eqnarray}
    3H\left(1+Q\right)\dot{\phi}=-V^{\prime}(\phi),\label{EvolInfQ}\\ 
    \quad \rho_R=\frac{3}{4}Q\Dot{\phi}^2\label{EvolDensitiesQ}
    \\ Q = C_\phi T^p \phi^q M^{1-p-q}/3H \label{QGen}
\end{eqnarray}
where $C_\phi$ is the dissipation coefficient and $Q$ is the generalized dissipation coefficient ratio\cite{universe9030124}, which is dimensionless. 
The power of temperature $T^p$ with $p=1$ corresponding to linear and $p=3$ as cubic dissipation regimes. In this study, we focus only on temperature dependence ignoring $\phi$ in the dissipation ratio, and hence we take $q=0$. The dissipation ratio $Q$ characterizes the efficiency of energy transfer from the inflaton field to the radiation bath. Its magnitude determines two distinct regimes of warm inflation: the \textit{weak dissipative regime} (WDWI) when $Q \ll 1$ and the \textit{strong dissipative regime} (SDWI) when $Q \gg 1$.



The slow roll parameters for warm inflation $\sigma_V$, $\eta_V$ \& $\epsilon_V$ are given by:
\begin{eqnarray}
\epsilon_V&=&\frac{m_p^2}{2}\left(\frac{V_{,\phi\phi}}{V}\right)^2,~~\eta_V=m_p^2\frac{V_{,\phi\phi}}{V},~~~ \sigma_V=\frac{m_p^2 V_{,\phi}}{\phi V}. \label{SlowRollParaWI} \\
\epsilon&=&\frac{\epsilon_V}{1+Q},\quad \eta=\frac{\eta_V-\epsilon_V}{1+Q}
\end{eqnarray}
 while the Hubble slow-roll parameters $\epsilon$ and $\eta$ explicitly depend on the dissipation ratio $Q$.

The standard power spectrum for warm inflation is given by 
\begin{eqnarray}
P_s &=& \left(\frac{H_*^2}{2 \pi m_p \dot{\phi} }\right)^2 \left(1+2 n_{BE} + \frac{T_*~~ 2\sqrt{3} \pi  Q_* }{H_* \sqrt{3+4 \pi  Q_*}}\right) G(Q_*),\\
n_{BE}&=&\frac{1}{\exp \left(\frac{H_*}{T_*}\right)-1},\\
G(Q) &=& 1+4.981 Q^{1.946}+0.127 Q^{4.33},\\
n_s &=& 1 + \frac{\partial\ln{P_s}}{\partial\ln{k}}=1 + (1-\epsilon)^{-1}\frac{\partial\ln{P_s}}{\partial\ln{N}}~.
\end{eqnarray}
We use this basic definition of spectra where $G(Q)$ is called the growth factor and $n_{BE}$ the Boltzmann statistics of the photon bath.

\section{Numerical Analysis}\label{Numerical}
We perform a detailed numerical study of the warm hybrid axion inflation dynamics and compute the resulting primordial power spectra. The background evolution is governed by the coupled inflaton–radiation system in the presence of a dissipative term $\Upsilon(T)$ arising from axion–gauge field interactions. To solve the system efficiently, we adopt dimensionless variables and use the number of e-folds $N$ as the time variable \cite{Benetti:2016jhf}:
\begin{eqnarray}
x &=& \phi/m_p, \quad y = \dot{\phi}, \quad z = m_p H, \quad r = \rho_r, \\
x_{,N} &=& y/z, \label{eq:dxdN} \\
y_{,N} &=& -3(1+Q)y - V_{,x}/z, \label{eq:dydN} \\
r_{,N} &=& -4 r + 3 Q y^2, \label{eq:drdN} \\
z^2 &=& \frac{1}{3}\left( \frac{y^2}{2} + V(x) + r \right), \label{eq:z2}
\end{eqnarray}
where $Q \equiv \Upsilon/(3H)$ is the dissipative ratio and $V(x)$ is the axion potential. The subscript $(_{,N})$ denotes a derivative with respect to $N$. We consider two temperature-dependent forms of the dissipation coefficient:
\begin{itemize}
    \item \textbf{Linear dissipation:} $\Upsilon(T) \propto T$,
    \item \textbf{Cubic dissipation:} $\Upsilon(T) \propto T^3$,
\end{itemize}
which correspond to distinct microscopic regimes of axion–gauge field interactions \cite{Berghaus:2020ekh}.

All numerical solutions are obtained using the \texttt{WI2easy} Mathematica code \cite{Rodrigues:2025neh}, a dedicated solver for warm inflation dynamics that integrates the background equations (\ref{eq:dxdN}--\ref{eq:z2}) together with the perturbation equations for scalar and tensor modes. The scalar power spectrum $P_s$ and tensor power spectrum $P_t$ are computed using the generalized warm inflation formalism \cite{Bastero-Gil:2016qru, Moss:2008yb}. The scalar spectral index $n_s$ and tensor-to-scalar ratio $r$ are evaluated at the pivot scale $k_* = 0.05\ \mathrm{Mpc}^{-1}$. We calibrate the potential amplitude using the measured value of $P_s$ from Planck  \cite{Planck:2018jri}, which also fixes $T^*/H^*$ for each benchmark point. We have verified the consistency of our \texttt{WI2easy} results with other warm inflation solvers such as \texttt{WI-easy} \cite{Rodrigues:2025neh} and ensured numerical convergence across the parameter space.

\begin{figure}[ht]
    \centering
        \includegraphics[width=0.48\textwidth]{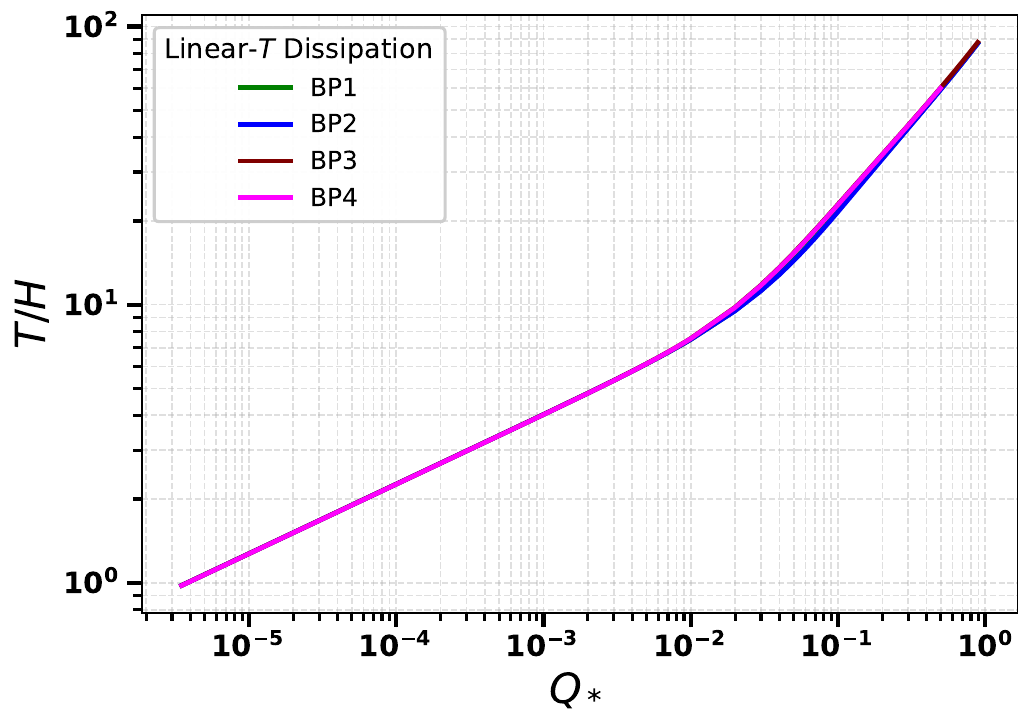}
    \includegraphics[width=0.48\textwidth]{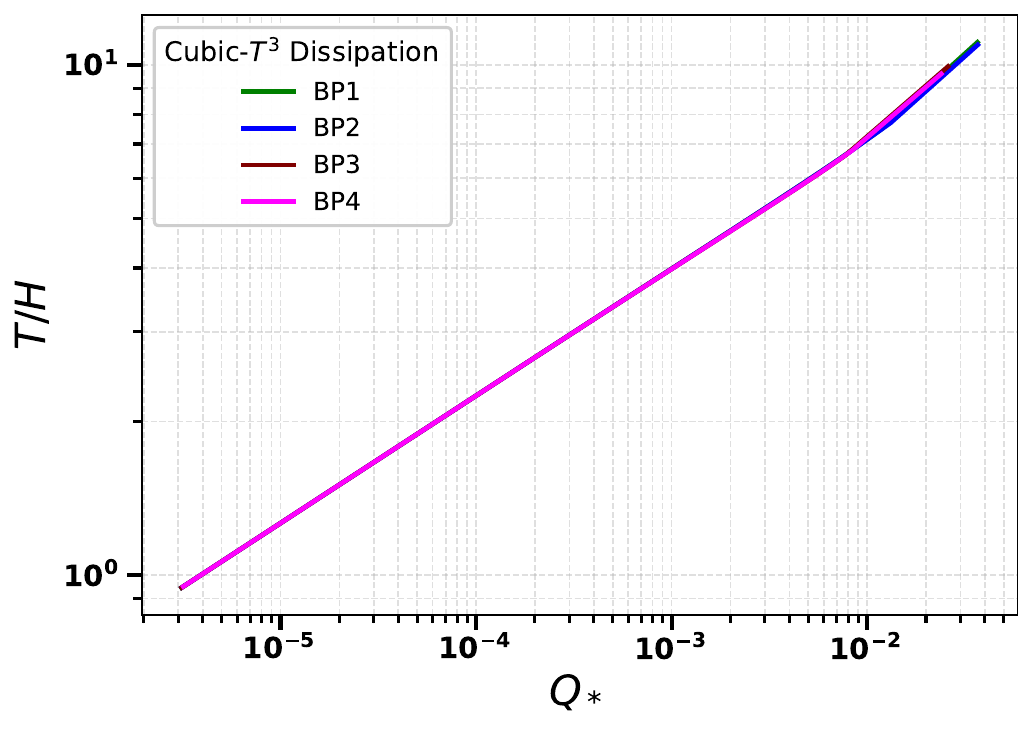}
    \includegraphics[width=0.48\textwidth]{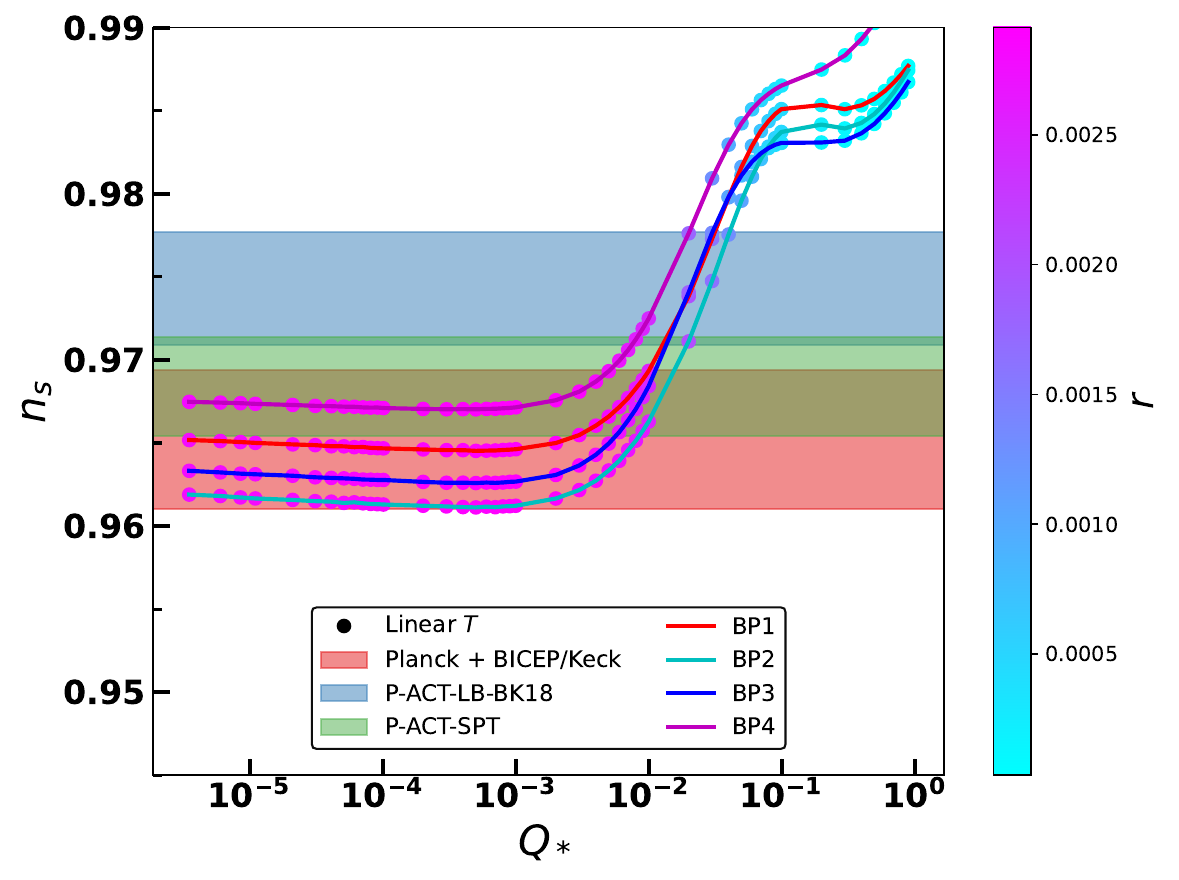}
    \includegraphics[width=0.48\textwidth]{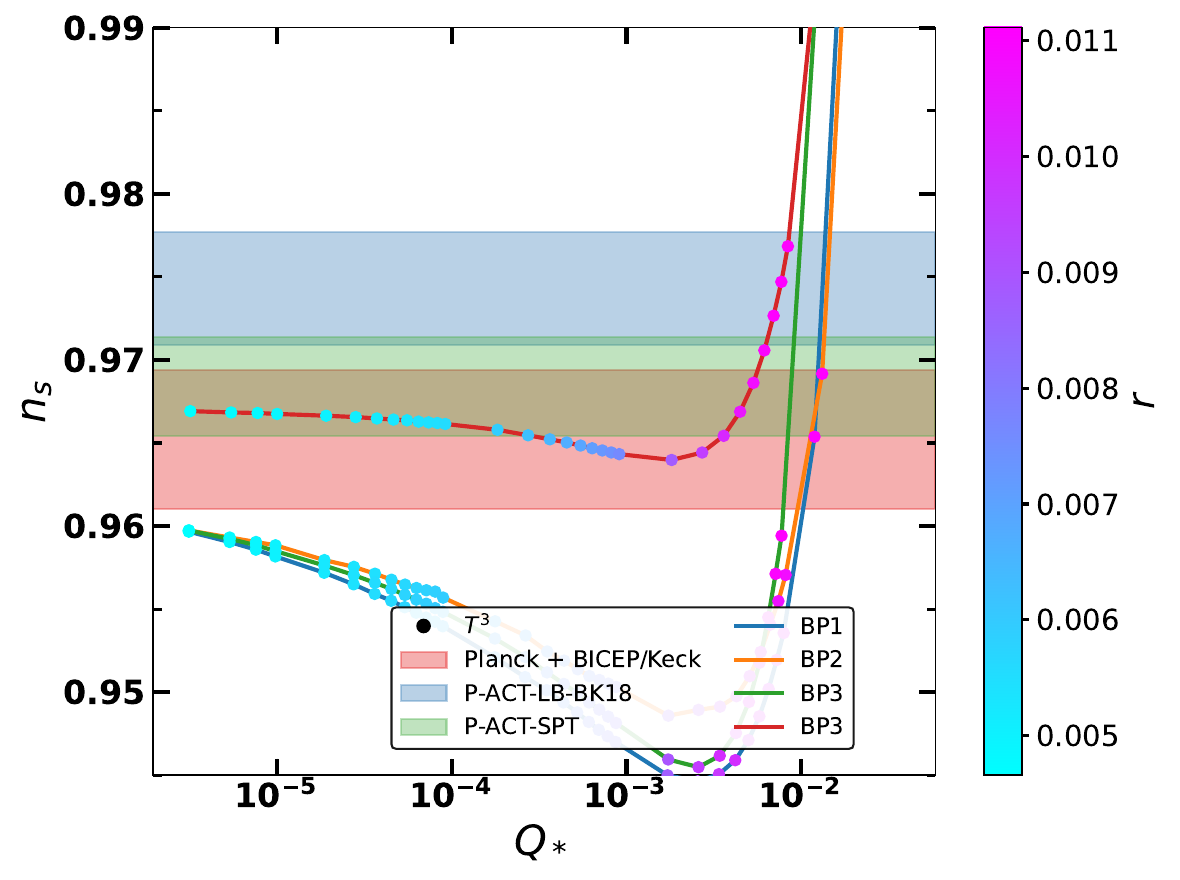}
    \caption{The top panel shows the relationship between the dissipation ratio $Q_*$ and the thermal ratio $T_*/H_*$ for both linear ($\Upsilon \propto T$) and cubic ($\Upsilon \propto T^3$) dissipation regimes, whereas the bottom panel shows the scalar spectral index $n_s$ as a function of $Q_*$ for linear dissipation (left) and cubic dissipation (right). The color bands represent the observational constraints from \emph{Planck+BICEP/Keck}, P-ACT-LB-BK18, and SPT datas. The four benchmark points (BP1--BP4) are indicated as colored trajectories.}
    \label{fig:ns_vs_Q}
\end{figure}

To systematically explore the model predictions, we select four benchmark points (BP1--BP4) that span the weak, intermediate, and strong dissipative regimes. Their parameters are listed in Table~\ref{tab:benchmarks}. 
\begin{table}[ht]
    \centering
    \caption{Parameters for the four benchmark points used in the numerical scan. }
    \begin{tabular}{c c c c c c }
        \hline
        Benchmark & $V_0$ & $m/m_p$ & $f/m_p$  & $V0_{Linear}$ &   $V0_{Cube}$  \\
        \hline
        BP1 & $V0\times 10^{-10}$ & $\sqrt{V0}\times2.8\times10^{-6}$ & $100$& $0.015$& $0.021$ \\
        BP2 & $V0\times 10^{-8}$ & $\sqrt{V0}\times2.8\times10^{-6}$ & $100$ & $0.0086$& $0.0093$\\
        BP3 & $V0\times 10^{-10}$ & $\sqrt{V0}\times2.8\times10^{-6}$ & $10$& $0.0019$& $0.384$ \\
        BP4 & $V0\times 10^{-10}$ & $\sqrt{V0}\times2.77\times10^{-6}$ & $1$& $0.782$& $0.813$ \\
        \hline
    \end{tabular}
    \label{tab:benchmarks}
\end{table}
The main results are displayed in Fig.~\ref{fig:ns_vs_Q} top panel illustrates the relationship between the dissipation ratio $Q_*$ and the thermal ratio $T_*/H_*$ across the full parameter space, revealing a transition at $Q_* \simeq 3.5\times10^{-5}$. In the cold inflation regime ($Q_* < 3.5 \times 10^{-5}$), $T_*/H_* < 0.1$, implying that the thermal bath is negligible compared to the Hubble scale, and quantum fluctuations dominate.

In the transition region ($3.5\times 10^{-5} < Q_* $), $0.1 < T_*/H_* < 1$, indicating that thermal effects are comparable to quantum effects. For the warm inflation regime ($Q_* > 10^{-5}$), $T_*/H_* > 1$, so thermal fluctuations dominate over quantum fluctuations and significantly affect the inflaton dynamics. 
The thermal transition behavior differs between linear ($\Upsilon \propto T$) and cubic ($\Upsilon \propto T^3$) dissipation forms. For linear dissipation, the transition is sharper at $Q_* \simeq 2.8 \times 10^{-5}$, thermalization occurs faster due to linear temperature dependence, and the correlation between the thermal ratio and dissipation is $T_*/H_* \propto Q_*^{0.85}$ for $Q_* > 10^{-4}$. In contrast, cubic dissipation shows a broader transition centered at $Q_* \simeq 4.2 \times 10^{-5}$, with slower initial thermalization but stronger temperature feedback, giving $T_*/H_* \propto Q_*^{0.45}$ for $Q_* > 10^{-4}$.

The bottom panel shows in Fig.~\ref{fig:ns_vs_Q}, which shows the scalar spectral index $n_s$ as a function of the dissipative ratio at horizon crossing $Q_*$ for both linear (left panel) and cubic (right panel) dissipation cases. The colored curves correspond to the four benchmark points, and the shaded regions represent the $68\%$ and $95\%$ confidence levels from combined CMB datasets (Planck+BK18, P-ACT-LB-BK18, SPT).

In the limit $Q_* \to 0$, the model reduces to cold inflation. For all benchmarks, $n_s$ asymptotically reaches a value $\approx(0.96-0.965)$, which may be in the preferred range of Planck $n_s = 0.9649 \pm 0.0042$ \cite{Planck:2018jri}. However, these results of cold inflation fall outside the combined Planck and ACT DR4/DR6 (TT+TE+EE) bounds in $n_s = 0.9743 \pm 0.01$. Thus, cold inflation is not best fit regime of our model, as it fails to match the observed spectral tilt of $n_s = 0.9743$ by P-ACT-LB-BK18.

As $Q_*$ increases, dissipation becomes dynamically important. Both dissipation prescriptions exhibit a monotonic decrease in $n_s$ with $Q_*$, but the rate of decrease is steeper for the cubic case due to the stronger temperature dependence. For linear dissipation, $n_s$ enters the $95\%$ CL region around $Q_* \sim 10^{-2}$, similarly for cubic dissipation this occurs already at $Q_* \sim 10^{-2}$. For $Q_* \gg 0.01$, both cases yield $n_s \approx 0.97$–$0.985$, which is in excellent agreement with the observations of P-ACT-LB-BK18 and SPT.

The tensor-to-scalar ratio $r$ is strongly suppressed in the warm regime; we find $r \sim 10^{-3}$--$10^{-2}$, which lies below the sensitivity thresholds of upcoming CMB experiments such as LiteBIRD and CMB-S4

All four benchmark points produce trajectories in the $n_s$–$Q_*$ plane that pass through the observationally allowed regions. This demonstrates the robustness of warm hybrid axion inflation: a wide range of dissipation strengths can yield predictions consistent with current CMB data. Notably, the cubic dissipation case allows for consistency at smaller $Q_*$, reducing the required dissipation strength for agreement with Planck.

\begin{figure}[ht]
    \centering
    \includegraphics[width=0.48\textwidth]{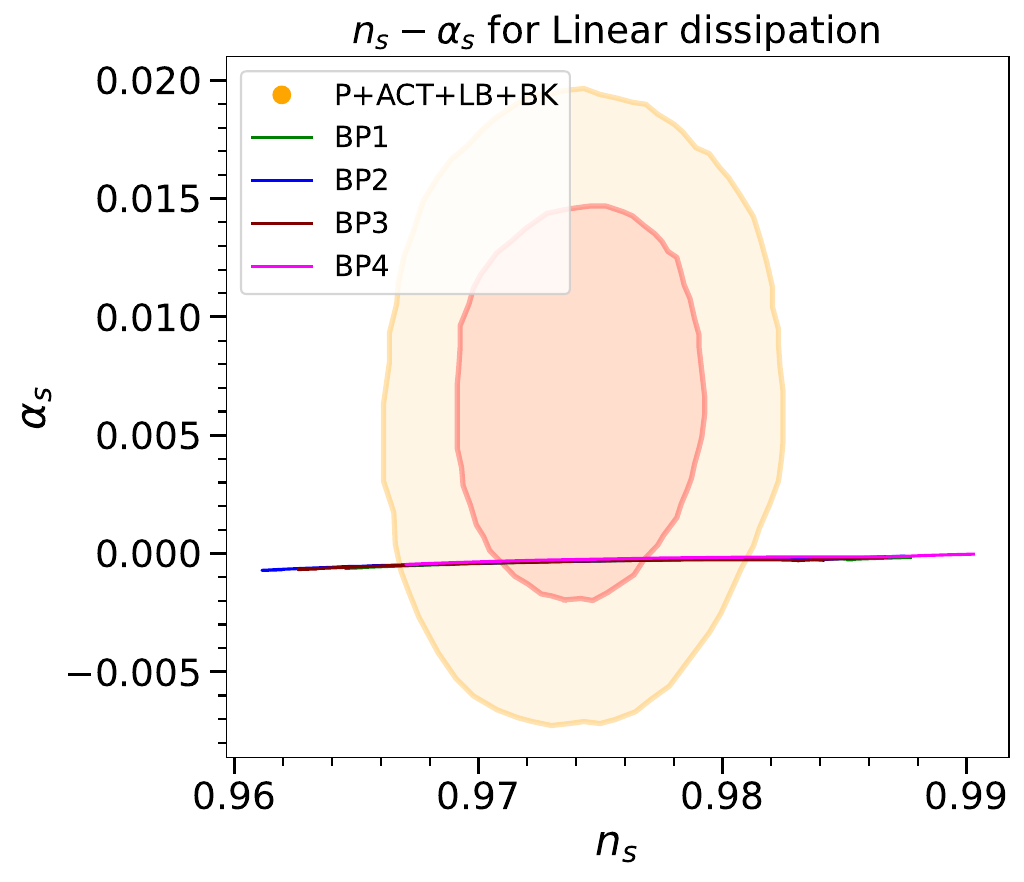}
    \includegraphics[width=0.48\textwidth]{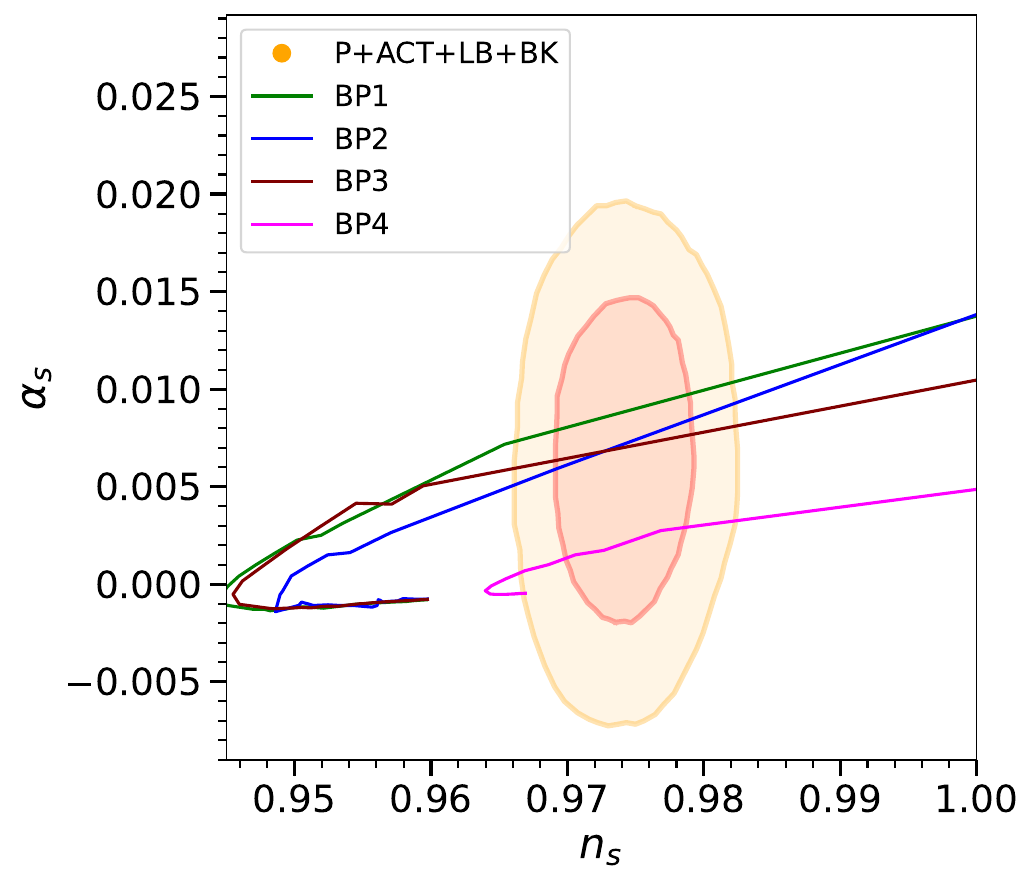}
   \caption{
        Predictions for the $n_s$ versus running of the scalar spectral index $\alpha_s$ for the four benchmark points 
        (BP1--BP4) in linear dissipation (left) and cubic dissipation (right). The shaded regions correspond to the 
        68\% and 95\% confidence levels from combined CMB datasets: P-ACT-LB-BK18 .
    }
    \label{fig:ns_vs_alpha}
\end{figure}

Similarly in Fig.~\ref{fig:ns_vs_alpha}, we present the predictions for $n_s$  as a function of $\alpha_s$ for the same four benchmark points (BP1--BP4), with observational constraints overlaid from combined datasets (P-ACT-LB-BK18 ).

All benchmark points yield predictions within or near the 95\% confidence regions of current CMB data, which constrain $\alpha_s = -0.0041 \pm 0.0067$ \cite{Planck:2018jri}. The mild negative running predicted in the warm regime may be probed with future high-precision CMB and large-scale structure surveys offering a potential discriminant between cold and warm inflation scenarios.

\section{Conclusion}\label{sec:conclusion}

In this work, we have investigated warm hybrid axion inflation within the framework of $\alpha$-attractor models. We considered a two-field scenario where the inflaton is an axion-like particle with a non-canonical kinetic term, and the waterfall field is responsible for ending inflation. The warm inflation dynamics are driven by dissipative effects, characterized by a dissipation coefficient $\Upsilon$, which couples the inflaton to a thermal bath. We considered two forms of the dissipation coefficient: linear ($\Upsilon \propto T$) and cubic ($\Upsilon \propto T^3$) in temperature.

Our numerical analysis, performed using the \texttt{W2easy} code, revealed that the model transitions from a cold inflation regime (disfavored by current data) to a warm inflation regime that is consistent with the latest CMB observations from Planck+BICEP/Keck, P-ACT-LB-BK18, and SPT. In particular, we found that for dissipative ratios $Q_* \gtrsim 0.01$, the scalar spectral index $n_s$ falls within the 95\% confidence region of the combined P-ACT-LB-BK18 data. The tensor-to-scalar ratio $r$ is strongly  in the warm regime, typically below $10^{-2}-10^{-3}$, which is with in the sensitivity of upcoming CMB experiments. Furthermore, the running of the scalar spectral index $\alpha_s$ in the strong dissipative regime also consistent with P-ACT-LB-BK18, which may be probed by future observations.

\section*{Acknowledgment}

The authors would like to thank Rudnei O. Ramos and Anish Ghoshal for their valuable discussions and constructive insights that contributed to this work. Special appreciation is extended to Rudnei O. Ramos for his assistance in clarifying the \textit{WI2easy} framework and for his guidance on the implementation of the proposed model.

\bibliographystyle{unsrt}

\bibliography{main.bib}

\end{document}